\documentclass[a4paper,12pt]{article}
\usepackage{amssymb,amsmath,amsfonts}
\usepackage{fullpage}
\begin{document}
\title{Skewness of maximum likelihood estimators in dispersion models}
\author{Alexandre B. Simas$^{a,}$\footnote{E-mail: alesimas@impa.br},~ Gauss M. Cordeiro$^{b,}$\footnote{E-mail: gausscordeiro@uol.com.br}~ and Andr\' ea V. Rocha$^{c,}$\footnote{E-mail: andrea@de.ufpb.br}\\\\
\centerline{\small{
$^a$Associa\c{c}\~ao Instituto Nacional de Matem\'atica Pura e Aplicada, IMPA,}}\\
\centerline{\small{
Estrada D. Castorina, 110, Jd. Bot\^anico, 22460-320, Rio de Janeiro-RJ, Brasil}}\\
\centerline{\small{
$^b$Departamento de Estat\' \i stica e Inform\' atica,
Universidade Federal Rural de Pernambuco,
}}\\
\centerline{\small{Rua Dom Manoel de Medeiros s/n, Dois Irm\~ aos,
52171-900 Recife-PE, Brasil}}\\
\centerline{\small{$^c$Departamento de Estat\' \i stica,
Universidade Federal da Para\' iba,
}}\\
\centerline{\small{
Cidade Universit\' aria - Campus I, 58051-970, Jo\~ ao Pessoa-PB, Brasil}}}

\date{}
\maketitle
\sloppy
\begin{abstract}
We introduce the dispersion models with a regression structure to extend the
generalized linear models, the exponential family nonlinear models
(Cordeiro and Paula, 1989) and the proper dispersion models (J\o rgensen, 1997a).
We provide a matrix expression for the skewness of the maximum likelihood estimators
of the regression parameters in dispersion models. The formula is suitable for
computer implementation and can be applied for several important submodels
discussed in the literature. Expressions for the skewness of the maximum
likelihood estimators of the precision and dispersion parameters are also derived.
In particular, our results extend previous formulas obtained by Cordeiro and
Cordeiro (2001) and Cavalcanti et al. (2009). A simulation study is 
perfomed to show the practice importance of our results.\\\\
\emph{Keywords}: dispersion models; nonlinear models; skewness; maximum likelihood.
\end{abstract}

\section{Introduction}

The assumption of symmetry plays a crucial role in many statistical procedures.
The notion of skewness of a distribution is related to a symmetry property.
The most commonly used measure of skewness is the standardized third cumulant defined by
$\gamma_1=\kappa_3/\kappa_2^{3/2}$, where $\kappa_r$ is the $r$th cumulant
of the distribution. In fact, the classical tests of symmetry use the standardized
third sample cumulant measure. A departure from the normal value of zero then indicates skewness.
Intuitively, we think of a distribution as being skewed if it systematically deviates from symmetry
by leaning to one side. Clearly, if the distribution is symmetrical, $\gamma_1$ vanishes
and therefore its value will give some indication of the extent of departure from symmetry.
However, there are asymmetrical distributions with as many zero-odd order central moments
as desired, so the value of $\gamma_1$ must be interpreted with some caution.
When $\gamma_1>0$ ($\gamma_1<0$), the distribution is positively (negatively) skewed and
will have a longer (shorter) right tail and a shorter (longer) left tail.

The value of the index $\gamma_1$ has been suggested as a possible measure of
non-normality of the distribution. We are concerned with the asymptotic skewness of
the distribution of the maximum likelihood estimators (MLEs) in the class of
dispersion models (DMs) (J\o rgensen, 1997b). This class of models represents a collection
of probability density functions that contains as sub-models: the proper dispersion
models (PDMs) (J\o rgensen, 1997a) and the well-known one-parameter exponential
families.

We assume that the random variables $Y_1,\ldots,Y_n$ are independent
and each $Y_i$ has a probability density function (pdf) of the form
\begin{equation}\label{density}
\pi(y;\mu_i,\phi)=\exp\{\phi t(y,\mu_i)+a(\phi,y)\},\quad y\in\mathbb{R},
\end{equation}
where $t(\cdot,\cdot)$ and $a(\cdot,\cdot)$ are known functions, $\phi>0$ and $\mu$ varies in
an interval of the real line. If $Y$ is continuous, $\pi$ is assumed to be a density with respect
to Lebesgue measure, while if $Y$ is discrete, $\pi$ is assumed to be a density with respect to
counting measure. We call $\phi$ the precision parameter and $\sigma^2 = \phi^{-1}$
the dispersion parameter. Similarly, the parameter $\mu$ may generally be
interpreted as a kind of location parameter not necessarily the mean of the distribution.
In practice, certain simplifications may be desirable. Exponential dispersion
models (EDMs) represent a special case of DMs for $t(y,\mu)=\theta y-b(\theta)$,
where $\mu=b'(\theta)$. The PDMs are also a special case of (\ref{density}) for
$a(\phi,y)= d_1(\phi)+d_2(y)$, where $d_1(\cdot)$ and $d_2(\cdot)$ are known functions.

We introduce a regression structure to (\ref{density})
\begin{equation}\label{reg}
h(\mu_i)=\eta_i=f(x_i;\beta),
\end{equation}
where $x_i = (x_{i1},\ldots,x_{im})^T$ is an $m$-vector of non-stochastic independent
variables associa\-ted with the $i$th response, $\beta=(\beta_1,\ldots,\beta_p)^T$ is
a $p$-vector of unknown parameters, $h(\cdot)$ is a known one-to-one twice continuously
differentiable function, usually referred to as the link function, and $f(\cdot;\cdot)$
is a possibly nonlinear, twice continuously differentiable function with respect to $\beta$.
The regression structure relates the covariates $x_i$ to the parameter of
interest $\mu_i$. The $n\times p$ matrix of derivatives of $\eta$ with respect
to $\beta$, specified by $\widetilde{X}=\widetilde{X}(\beta)=\partial\eta/\partial\beta$,
is assumed to have rank $p$ for all $\beta$. The DM defined by
equations (\ref{density}) and (\ref{reg}) is a general model that allows
for parsimonious representation. We assume that the usual regularity conditions for
maximum likelihood estimation and large sample inference hold;
see Cox and Hinkley (1974, Chapter 9).

From now on, the term ``dispersion model'' (denoted simply by DM)
represents a regression model specified by (\ref{density})
and (\ref{reg}) that allows for parsimonious representation.
For DMs, Rocha et al. (2009) obtained a matrix expression for the
covariance matrix of the MLEs up to order $O(n^{-2})$, where $n$ is the
sample size, Simas et al. (2009a) calculated the second-order biases of
the estimators of the parameters and Simas et al. (2009b) studied
asymptotic tail properties for some distributions belonging to the class of dispersion models.

The DMs extend the exponential family nonlinear models (EFNLMs) (Cordeiro
and Paula, 1987), since they contain many distributions that are not
in the exponential family form, whereas the EFNLMs generalize the well-known
generalized linear models (GLMs), since they allow a nonlinear regression
structure. Paula (1992) derived general expressions for the second-order
biases of the MLEs in EFNLMs, thus extending previous result by
Cordeiro and McCullagh (1991) for GLMs. Wei (2004) wrote an excellent
book on these models. More recently, Simas and Cordeiro (2009) proposed corrected 
Pearson residuals in EFNLMs and Simas et al. (2009a) proposed corrected MLEs in DMs,
thus extending the results by Cordeiro and McCullagh (1991) and Paula (1992).

The PDMs contain several important non-exponential models, for instance,
the von-Mises regression model for data distributed along the unit circle
and the simplex model for data distributed in the standard unit
interval $(0,1)$. A complete study of PDMs is presented by J\o rgensen (1997b).

Few attempts have been made to develop second-order asymptotic theory for DMs
in order to have better likelihood inference procedures. An asymptotic formula of order
$n^{-1/2}$ for the skewness of the distribution of $\hat{\beta}$ in GLMs was derived by
Cordeiro and Cordeiro (2001). In this article, we provide asymptotic
formulae for the third cumulants of the distributions of the MLEs of the
regression parameters $\beta$, precision parameter $\phi$ and dispersion
parameter $\sigma^2$ in DMs thus extending the results by Cordeiro and Cordeiro (2001).
The formulae are useful to define the skewness of these distributions
corrected to order $n^{-1/2}$. The knowledge of the skewness can be used as a measure of
departure of these distributions from normality. We consider asymptotic results for likelihood
inference with respect to the vector $\beta$ of parameters and
scalars $\phi$ and $\sigma^2$ for large $n$.

The rest of the paper is organized as follows. In Section 2, we apply the general formula for
the third cumulant of the MLE given by Bowman and Shenton (1998) to obtain a simple
expression for the skewness of the distribution of the MLE $\hat{\beta}$. Section 3
is devoted to the skewness of the distributions of the MLEs $\hat{\phi}$ and
$\hat{\sigma}^2$. In Section 4, we apply our main result to a number of important
special models. In Section 5, we provide simulation results for the reciprocal gamma nonlinear
model to investigate the skewness of the MLEs in DMs and to motivate the use of the proposed
formula. Some concluding remarks are given in Section 6.

\section{Skewness of $\hat\beta$}

In this section, we derive the skewness of the MLEs of the
parameters $\beta$ in DMs. Consider the observations $y_1,\ldots,y_n$ and
let $\ell=\ell(\beta,\phi)$ be the total log-likelihood function for $\beta$
and $\phi$. We assume that the usual regularity conditions for maximum likelihood estimation
and large sample inference hold (Cox and Hinkley, 1974, Chapter 9). A simple calculation
shows that $E(\partial^2\ell/\partial\phi\partial\beta)=0$, and then the parameters
$\beta$ and $\phi$ are globally orthogonal (Cox and Reid, 1987). Let $\hat{\beta}$
and $\widehat{\phi}$ be the MLEs of $\beta$ and $\phi$, respectively, and $\mu_i = h^{-1}(\eta_i)$
be the inverse link function. Then, the unit deviance for the DM, given the
data vector $y$, is defined by
$$D(y,\mu) = 2\sum_{i=1}^n [\sup_{\mu}t(y_i,\mu) - t(y_i,\mu_i)].$$
The MLE of $\beta$ can be calculated by minimizing the deviance $D(y,\mu)$
with respect to $\beta$. The maximum likelihood equations for $\beta$
do not depend on the precision parameter $\phi$ and are given
by $\widetilde{X}^T t'(y,\mu)=0$, where $t'(y,\mu)=\partial t(y,\mu)/\partial\mu$
is an $n \times 1$ vector. These nonlinear equations have the same form of the standard
estimating equations for GLMs and can be solved by iterative methods. Alternatively,
we can maximize directly minus the deviance $-D(y,\mu)$, for example, using some
standard statistical software such as SAS or the GAMLSS package in R.

Given the estimate $\hat{\beta}$, the MLE of $\phi$ is obtained as the solution
of the nonlinear equation
$$\sum_{i=1}^n a'(y_i,\phi) = \frac{1}{2} D(y,\widehat{\mu}) - \sum_{i=1}^n \sup_{\mu} t(y_i,\mu),$$
where $a'(\phi,y)=\partial a(\phi,y)/\partial\phi$. The MLE $\hat{\phi}$ of the
precision parameter is a function of the deviance of the model. The MLE of the dispersion
parameter $\sigma^2$ is $\hat{\sigma}^2=\hat{\phi}^{-1}$.

We define $d_r=d_r(\mu,\phi)=E\{\partial^r t(Y,\mu)/\partial\mu^r\}$ for $r=1,2,3$.
From some regularity conditions, we have $d_1=0$ and $d_2=-\phi E\{[\partial t(Y,\mu)/\partial\mu]\}^2$.
We shall use the following notation for the derivatives of the log likelihood
function $\ell=\ell(\beta,\phi)$: $\kappa_{rs}$ $=$ $E(\partial^2\ell/\partial\beta_r\partial\beta_s)$,
$\kappa_{rst}=E(\partial^3\ell/\partial\beta_r\partial\beta_s\partial\beta_t),$
$\kappa_{r,s}=E(\ell/\partial\beta_r\,\partial\ell/\partial\beta_s)$,
$\kappa_{r,s,t}=E(\partial\ell/\partial\beta_r\,\partial\ell/\partial\beta_s\,\partial\ell/\partial\beta_t),$
$\kappa_{r,st}=E(\partial\ell/\partial\beta_r\,\partial^2\ell/\partial\beta_s\partial\beta_t)$, etc.
Note that $\kappa_{r,s}= -\kappa_{rs}$ and that $\kappa_{rs,t}$ is the covariance of the
first derivative of $\ell$ with respect to $\beta_t$ with the mixed second derivative with
respect to $\beta_r$ and $\beta_s$. All $\kappa$'s refer to a total over the sample and
are, in ge\-neral, of order $n$. The total Fisher information matrix has
elements $\kappa_{r,s}= -\kappa_{rs}$ and let $\kappa^{r,s}$ be the corresponding elements
of its inverse. The joint information matrix for $\gamma=(\beta^T,\phi)^T$
is $K_\gamma={\rm diag}\{\phi\widetilde{X}^T W\widetilde{X},na^{(2)}\}$, where
$W = {\rm diag}\{-d_2 (d\mu/d\eta)^2\}$ and $a^{(2)}=a^{(2)}(\mu,\phi)=-E\{\partial^2 a(\phi,Y)/\partial\phi^2\}$.
The MLEs of $\beta$ and $\phi$ are asymptotically independent due to their asymptotic normality and
the block diagonal structure of the joint information matrix $K_\gamma$.

We introduce the notation $(r)_i = \partial\eta_i/\partial\beta_r$, $(r,s)_i = (\partial\eta_i/\partial\beta_r)(\partial\eta_i/\partial\beta_s)$,
$(r,st)_i = (\partial\eta_i/\partial\beta_r)(\partial^2\eta_i/\partial\beta_s\partial\beta_t)$, etc.
Let $\kappa_3(\hat{\beta}_a)=E\{(\hat{\beta}_a-\beta_a)^3\}$ be the third cumulant of the
MLE $\hat{\beta}_a$ of $\beta_a$ for $a=1,\ldots,p$. From the general expression for the
multi-parameter $n^{-2}$ third cumulants of the MLEs given by Bowman and Shenton (1998),
we can write to order $n^{-2}$
\begin{equation}\label{skew}
\kappa_3(\hat{\beta}_a)={\sum}'\kappa^{a,r}\kappa^{a,s}\kappa^{a,t}(\kappa_{r,s,t}+3\kappa_{rst}+6\kappa_{rs,t}).
\end{equation}
In equation (\ref{skew}), $\sum'$ denotes the summation over all $p+1$ parameters $\beta_1,\ldots,\beta_p$
and $\phi$. Let $\Sigma$ be the summation over the observations. The key for obtaining a simple
expression for $\kappa_3(\hat{\beta}_a)$ in DMs is the invariance of the $\kappa$'s under permutation
of parameters $\beta'$s and the orthogonality between $\phi$ and $\beta$ (Cox and Reid, 1987),
i.e., $E(-\partial^2\ell/\partial\beta\partial\phi)=0$.

After some calculation and using the notation of Cordeiro et al. (1994), we obtain
$$\kappa_{rst}= -\phi \sum_{i=1}^n [(f+2g)_i (r,s,t)_i + w_i \{(r,st)_i + (s,rt)_i + (t,rs)_i\}],$$
$$\kappa_{r,st}=\phi \sum_{i=1}^n [(f-e)_i (r,s,t)_i + w_i (r,st)_i]\hbox{~~and~~}
\kappa_{r,s,t}= -\phi\sum_{i=1}^n [(2f-2g-3e)_i (r,s,t)_i],$$
where
$$f=-\frac{d\mu}{d\eta}\frac{d^2\mu}{d\eta^2}d_2-\left(\frac{d\mu}{d\eta}\right)^3 d_3,\quad g=-\frac{d\mu}{d\eta}\frac{d^2\mu}{d\eta^2}d_2,$$
$$e= -\left(\frac{d\mu}{d\eta}\right)^3 d_2'\hbox{~~~~and~~~~} w=-\left(\frac{d\mu}{d\eta}\right)^2 d_2,$$
where $d_2'$ is the first partial derivative of $d_2$ with respect to $\mu$. Because of the
orthogonality between $\phi$ and $\beta$, we have only to take into account in equation (\ref{skew})
the sum of terms involving the various combinations of the parameters $\beta$. Hence, the crucial
quantity $\kappa_{r,s,t} + 3\kappa_{rst}+6\kappa_{rs,t}$ for the $n^{-2}$ third central moment
of $\hat{\beta}_a$ is given by
\begin{equation}\label{result1}
\kappa_{r,s,t} + 3\kappa_{rst}+6\kappa_{rs,t} = \phi\sum_{l=1}^n [(f-4g-3e)_i(r,s,t)_i + 3w_i\{(r,st)_i-(s,rt)_i-(t,rs)_i\}].
\end{equation}
Inserting (\ref{result1}) in (\ref{skew}), inverting the order of the summation and
rearranging, we obtain
$$\kappa_3(\hat{\beta}_a)=\phi\sum_{l=1}^n(f-4g-3e)_i\left(\sum_{r=1}^p \kappa^{a,r}(r)_i\right)^3
-3\phi \sum_{l=1}^n w_i\left(\sum_{r=1}^p \kappa^{a,r} (r)_i\right)\left(\sum_{s,t=1}^p \kappa^{a,s}\kappa^{a,t} (st)_i\right).$$
Let $\phi K_\beta$ be the information matrix for $\beta$, where
$K_\beta = \widetilde{X}^T W\widetilde{X}$. Also, let $\rho_a^T$ and $\delta_i$ be $1\times p$
and $n\times 1$ vectors of zeros with one in the $a$th and $i$th components,
respectively. Thus, $\sum_{r=1}^p \kappa^{a,r} (r)_i=\phi^{-1}\rho_a^T K_\beta^{-1}\widetilde{X}^T\delta_i$.
Further, let $\widetilde{X}_i$ be a $p\times p$
matrix with elements $\partial^2\eta_i/\partial\beta_r\partial\beta_s$. Then,
$\sum_{s,t=1}^p \kappa^{a,s}\kappa^{a,t}(st)_i=\phi^{-2}\rho_a^T K_\beta^{-1} \widetilde{X}_i K_\beta^{-1}\rho_a$.
We define the matrices of order $p\times n$: $M=\{m_{al}\}=K_\beta^{-1}\widetilde{X}^T$ and
$N=\{n_{ai}\}=\{\rho_a^T K_\beta^{-1} \widetilde{X}_i K_\beta^{-1}\rho_a\}$.
The $O(n^{-2})$ third cumulant of $\hat{\beta}_a$ is
$$\kappa_3(\hat{\beta}_a)=\phi \sum_{i=1}^n (f-4g-3e)_i \frac{m_{ai}^3}{\phi^3}-3\phi\sum_{i=1}^n w_i \frac{m_{ai}n_{ai}}{\phi^3},$$
where $m_{ai}$ is the $(a,i)$th element of the matrix $M$. Let $\kappa_3(\hat{\beta})=(\kappa_3(\hat{\beta}_1),\ldots,\kappa_3(\hat{\beta}_p))^T$ be the
$p\times 1$ vector of the $n^{-2}$ third cumulants of the $\hat{\beta}$'s.
The third cumulant vector has a simple expression
\begin{equation}\label{secskew}
\kappa_3(\hat{\beta})=\frac{1}{\phi^2}\{M^{(3)}(f-4g-3e)- 3(M\odot N)w\},
\end{equation}
where $f=(f_1,\ldots,f_n)^T, g=(g_1,\ldots,g_n)^T, e=(e_1,\ldots,e_n)^T$ and $w=(w_1,\ldots,w_n)^T$
are $n\times 1$ vectors, whose elements were previously defined, $M^{(3)} = M\odot M\odot M$,
and $\odot$ is the Hadamard (direct) product. Expression (\ref{secskew}) is a
function of the model matrix $\widetilde{X}$, the matrices $\widetilde{X}_i$
for $i=1,\ldots,n$, the first three derivatives of the function $t(\cdot,\cdot)$
with respect to $\mu$ and the unknown $\mu$'s. The third cumulant vector is easily
computed since it involves only simple operations on matrices and vectors.
The vector $\kappa_3(\hat{\beta})$ is weighted by the inverse of the square
of the precision parameter. Equation (\ref{secskew}) generalizes previous results
obtained by Cordeiro and Cordeiro (2001) and Cavalcanti et al. (2009) for GLMs
and EFNLMs, respectively.

From the third cumulant vector (\ref{secskew}) and the asymptotic covariance
matrix Cov$(\hat{\beta})=\phi^{-1}(\widetilde{X}^T W\widetilde{X})^{-1}$ of $\hat{\beta}$,
we can easily obtain the asymptotic skewness $\gamma_1(\hat{\beta}_a)=\kappa_3(\hat{\beta}_a)/{\rm Var}(\hat{\beta}_a)^{3/2}$ of the distribution of the estimate $\hat{\beta}_a$ of
the regression parameter $\beta_a$ for $a=1,\ldots,p$. Clearly, $\gamma_1(\hat{\beta}_a)$
is of order $n^{-1/2}$ and is weighted by the inverse of the square root of the precision
parameter $\phi$. Thus, the normal approximation for the distribution of $\hat{\beta}$
deteriorates when $\phi$ decreases, which is consistent with the small dispersion asymptotics
phenomenon noted by J\o rgensen (1987b). The parameters $\phi$ and $\mu$ should be replaced by
consistent estimators $\hat{\phi}$ and $\widehat{\mu}$ to obtain a numerical value for $\widehat{\gamma}_1(\hat{\beta}_a)$. We can use the estimate of the skewness $\widehat{\gamma}_1(\hat{\beta}_a)$
as an indicator of departure from the normal distribution of $\hat{\beta}_a$.

By evaluating the skewness in (\ref{secskew}), we can obtain an
approximate Edgeworth expansion for the density function of the
estimate $\hat\beta_a$, whose leading terms are
$$f_{\hat\beta_a}(x)=\phi(x)\left\{1+\frac{\kappa_3(\widehat{\beta_a})}{6}H_3(x)+\frac{\kappa_3(\widehat{\beta_a})^2}{72}H_6(x)\right\},$$
where $\phi(x)$ is the standard normal density function and
$H_3(x)=x^3-3x$ and $H_6(x)=x^6-15x^4+45x^2-15$ are Hermite
polynomials, which should work better than the standard
normal distribution.

\section{Skewness of $\hat\phi$ and $\hat\sigma^2$}

We provide general formulae for the $n^{-2}$ third cumulants of the MLEs of the
precision and dispersion parameters in DMs. First, we consider the
third cumulant of the estimate $\hat{\phi}$ derived in Section 2 as a
solution of a nonlinear equation. Let $\alpha^{(r)}=E\{(\partial a(\phi, Y)/\phi)^r\}$ and
$\alpha_{r,s}=E\{\partial^r a(\phi,Y)/\partial\phi^r \partial^s a(\phi,Y)/\partial \phi^s\}$.
From the orthogonality between $\phi$ and $\beta$, equation (\ref{skew}) yields
$$\kappa_3(\hat{\phi})=\kappa^{\phi,\phi^3}(\kappa_{\phi,\phi,\phi}+ 3\kappa_{\phi\phi\phi}
+ 6\kappa_{\phi\phi,\phi}).$$
Let $\kappa_{\phi,\phi}=\alpha^{(2)}$, $\kappa_{\phi\phi,\phi}=\alpha_{2,1}$,
$\kappa_{\phi\phi\phi}=\alpha_{3,0}$ and $\kappa_{\phi,\phi,\phi}=\alpha^{(3)}$.
Thus, the third cumulant of $\hat{\phi}$ becomes
\begin{equation}\label{kappaphi}
\kappa_3(\hat{\phi})=\frac{\alpha^{(3)}+ 3\alpha_{3,0}+6\alpha_{2,1}}{[\alpha^{(2)}]^3}.
\end{equation}
From equation (\ref{kappaphi}) and the asymptotic variance Var$(\hat{\phi})=\left[\alpha^{(2)}\right]^{-1}$,
we obtain the asymptotic skewness of $\hat{\phi}$ as
$$\gamma_1(\hat{\phi})= \frac{\alpha^{(3)}+ 3\alpha_{3,0}+ 6 \alpha_{2,1}}{[\alpha^{(2)}]^{3/2}}.$$

We write (\ref{density}) in terms of $\sigma^2=\phi^{-1}$
\begin{equation}\label{densdisp}
\pi(y;\mu_i,\sigma^2)=\exp\{\sigma^{-2} t(y,\mu_i)+a_\ast(\sigma^2,y)\},\quad y\in\mathbb{R},
\end{equation}
where $a_\ast(\sigma^2,y)=a(\sigma^{-2},y)$. A straightforward calculation shows
that $\sigma^2$ and $\mu$ are orthogonal parameters. Let $\alpha_\ast^{r,s}=\partial^r E\{\partial^s a_\ast (\sigma^2,Y)/\partial(\sigma^2)^s\}/\partial(\sigma^2)^r$. We can obtain the cumulants
$\kappa_{\sigma^2,\sigma^2}=-2\sigma^{-2}\alpha_\ast^{0,1}-\alpha_\ast^{0,2}$,
$\kappa_{\sigma^2\sigma^2\sigma^2}= -6\sigma^{-4}\alpha_\ast^{0,1}+\alpha_\ast^{0,3}$, $\kappa_{\sigma^2\sigma^2,\sigma^2}=\alpha_\ast^{1,2} + 2\sigma^{-2}\alpha_\ast^{1,1}+4\sigma^{-4}\alpha_\ast^{0,1}
-\alpha_\ast^{0,3}$ and $\kappa_{\sigma^2,\sigma^2,\sigma^2} = -6\sigma^{-2} \alpha_\ast^{1,1}
-3\alpha_\ast^{1,2} - 6\sigma^{-4} \alpha_\ast^{0,1} + 2\alpha_\ast^{0,3}$. From
$$\kappa_{\sigma^2,\sigma^2,\sigma^2} + 3\kappa_{\sigma^2\sigma^2\sigma^2}+6\kappa_{\sigma^2\sigma^2,\sigma^2} = -6\sigma^{-2}\alpha_\ast^{1,1}+3\alpha_\ast^{1,2}-\alpha_\ast^{0,3},$$
we have
\begin{equation}\label{kappasigma2}
\kappa_3(\hat{\sigma}^2) = \frac{6\sigma^{-2}\alpha_\ast^{1,1} - 3\alpha_\ast^{1,2} + \alpha_\ast^{0,3}}{\left(2\sigma^{-2}\alpha_\ast^{0,1} + \alpha_\ast^{0,2}\right)^3}.
\end{equation}
From equation (\ref{kappasigma2}), the asymptotic skewness of the distribution
of $\hat{\sigma}^2$ is given by
$$\gamma_1(\hat{\sigma}^2) = \frac{-6\sigma^{-2}\alpha_\ast^{1,1} + 3\alpha_\ast^{1,2} - \alpha_\ast^{0,3}}{\left(-2\sigma^{-2}\alpha_\ast^{0,1} - \alpha_\ast^{0,2}\right)^{3/2}}.$$

\section{Some Special models}

Here, we examine some special cases of formulas (\ref{secskew}), (\ref{kappaphi})
and (\ref{kappasigma2}). Some other special cases could be easily derived
because of the advantage of the explicit matrix expression (\ref{secskew})
which is easily implemented in statistical packages or in a computer algebra
system such as Mathematica or Maple. Table \ref{tabelalink} lists the most common
link functions and the quantities required for the skewness of the MLE $\hat\beta$,
where $\Phi(\cdot)$ is the standard normal distribution function, $\phi(x)$ is the density of the standard normal
distribution and $\phi'(x)$ is its first derivative.

\begin{table}[hbt]
\caption{The most common link functions and their derivatives.}
\begin{center}
\begin{tabular}{lccc}
\hline
Link & Formula & ${d\mu}/{d\eta}$&${d^2\mu}/{d\eta^2}$ \\

\hline
Logit             &$\log\left({\mu}/{(1-\mu)}\right)=\eta$&$\mu(1-\mu)$&$\mu(1-\mu)(1-2\mu)$ \\
Probit            &$\Phi^{-1}(\mu)=\eta$&$\phi(\Phi^{-1}(\mu))$&$\phi'(\Phi^{-1}(\mu))$\\
Log               &$\log(\mu)=\eta$&$\mu$&$\mu$\\
Identity          &$\mu=\eta$&$1$&$0$\\
Reciprocal        &$\mu^{-1}=\eta$&$-\mu^2$&$2\mu^3$\\
Square reciprocal &$\mu^{-2}=\eta$&${-\mu^3}/{2}$&${3\mu^5}/{4}$\\
Square Root       &$\sqrt{\mu}=\eta$&$2\sqrt{\mu}$&$2$\\
C-loglog          &$\log(-\log(1-\mu))=\eta$&$-\log(1-\mu)(1-\mu)$&$-(1-\mu)\log(1-\mu)$\\
                  &                         &                     &$\times(1+\log(1-\mu))$\\
Tangent           &$\tan(\mu)=\eta$&$\cos(\mu)^2$&$2\cos(\mu)^3\sin(\mu)$\\
\hline
\end{tabular}
\end{center}
\label{tabelalink}
\end{table}

\subsection{Generalized Linear Models}

We calculate the skewness of the MLE $\hat\beta$. The function
$t(\cdot,\cdot)$ has the form $t(y,\theta)= y\theta-b(\theta)$,
where the mean value is $\mu=\tau(\theta)=b'(\theta)$ and the variance
function $V=V(\mu)$ is related to the mean by $d\tau^{-1}(\mu)/d\mu=V^{-1}$. We have $t\{y,\tau^{-1}(\mu)\}=y\tau^{-1}(\mu)-b\{\tau^{-1}(\mu)\}$.
For GLMs, $d_2=-V^{-1}$ and $d_3= 2V^{-2}V^{(1)}$,
where $V^{(1)}= dV(\mu)/d\mu$, $W=\{V^{-1}(d\mu/d\eta)^2\}$, $\widetilde{X}$
reduces to the matrix $X$, $h(\mu_i)=\eta_i=x_i^T \beta$ and $N$ vanishes.
From the matrix $M=\{m_{al}\}=(X^T W X)^{-1} X^T$ and by formula (\ref{secskew}),
we obtain
$$\kappa_3(\hat{\beta}_a)=\frac{1}{\phi^2} \sum_{i=1}^n m_{ai}^{3} \left\{3\frac{d\mu}{d\eta}\frac{d^2\mu}{d\eta^2}V^{-1}-2\left(\frac{d\mu}{d\eta}\right)^3 V^{-2}V^{(1)}\right\}_i,$$
which is identical to the result by Cordeiro and Cordeiro (2001).
Table \ref{tabelaexpf} lists the distributions in the exponential family
and the quantities required for the skewness.
\begin{table}
\caption{Expressions of $V$ and its derivatives for distributions in the exponential family.}
\begin{center}
\begin{tabular}{lccc}
\hline
Distribution&$V$&$V^{(1)}$&$V^{(2)}$\\
\hline
Normal      &$1$&$0$&$0$\\
Poisson     &$\mu$&$1$&$0$\\
Binomial    &$\mu(1-\mu)$&$1-2\mu$&$-2$\\
Gamma       &$\mu^2$&$2\mu$&$2$\\
Inver. Gaussian &$\mu^3$&$3\mu^2$&$6\mu$\\
\hline
\end{tabular}
\end{center}
\label{tabelaexpf}
\end{table}

We also calculate the skewness of the estimators of $\phi$ and $\sigma^2$
for two-parameter exponential family distributions with canonical
parameters $\phi$ and $\phi \theta$. We have $a(\phi,y)=\phi c(y)+a_1(\phi)+a_2(y)$,
where $c(\cdot)$ is a known function. We have $\alpha^{(2)}=-n a_1''(\phi)$, $\alpha_{3,0}= na_1'''(\phi)$,
$\alpha^{(3)}=-n a_1'''(\phi)$ and $\alpha_{2,1}=0$. From (\ref{kappaphi}), we obtain
$$\kappa_3(\hat{\phi})=-\frac{2 a_1'''(\phi)}{n^2 a_1''(\phi)^3},$$
and the skewness becomes
$$\gamma_1(\hat{\phi})= \frac{2 a_1'''(\phi)}{\sqrt{n}\{-a_1''(\phi)\}^{3/2}}.$$
These expressions agree with the results by Cordeiro and Cordeiro (2001).
We define $\xi(\sigma^2)=a_1(\sigma^{-2})$. From similar calculations,
and using (\ref{kappasigma2}), the second-order third cumulant of
$\hat{\sigma}^2$ can be expressed as
$$\kappa_3(\hat{\sigma}^2)=-\frac{2\sigma^4 \{\sigma^2\xi'''(\sigma^2)+3\xi''(\sigma^2)\}}{n^2\{2\xi'(\sigma^2)+\sigma^2\xi''(\sigma^2)\}^3},$$
which yields
$$\gamma_1(\hat{\sigma}^2) = \frac{2\sigma\{\sigma^2 \xi'''(\sigma^2)+3\xi''(\sigma^2)\}}{\sqrt{n}\{-2\xi'(\sigma^2)-\sigma^2\xi''(\sigma^2)\}^3}.$$

Table \ref{tabelaexpf2} lists the skewness of the MLEs of the parameters $\phi$
and $\sigma^2$. The function $a_1(\phi)$ is equal to $\log\sqrt{\phi}$, $\phi \log(\phi)-\log\Gamma(\phi)$
and $\log\sqrt{\phi}$ for the normal, gamma and inverse Gaussian distributions, respectively.
Here, $\Gamma(\cdot)$ is the gammma function and $\psi(\cdot)$ is the digamma function.
\begin{table}[!th]
\caption{Skewness of $\hat{\phi}$ and $\hat{\sigma}^2$.}
\begin{center}
\begin{tabular}{lcc}
\hline
Distribution&$\kappa_3(\hat{\phi})$&$\gamma_1(\hat\phi)$\\
\hline
Normal              &$\frac{16\phi^3}{n^2}$&$\frac{2^{5/2}}{\sqrt{n}}$\\
Gamma               &$\frac{2\phi(1+\phi^2\psi''(\phi)]}{n^2[1-\phi \psi'(\phi)]^3}$&
                     $\frac{-2[\psi''(\phi)+\phi^{-2}]}{\sqrt{n}[\psi'(\phi)-\phi^{-1}]^{3/2}}$\\
Inver. Gaussian         &$\frac{16\phi^3}{n^2}$&$\frac{2^{5/2}}{\sqrt{n}}$\\
\hline
&$\kappa_3(\hat{\sigma}^2)$&$\gamma_1(\hat{\sigma}^2)$\\
\hline
Normal              &$\frac{8\sigma^6}{n^2}$&$\frac{2^{3/2}}{\sqrt{n}}$\\
Gamma               &$\frac{-2\left[\frac{\psi''(\sigma^{-2})}{\sigma^{6}}+\frac{3\psi'(\sigma^{-2})}{\sigma^4}-\frac{2}{\sigma^2}\right]}
                     {n^2\left[\sigma^{-4}-\frac{\psi'(\sigma^{-2})}{\sigma^6}\right]^3}$&
                     $\frac{2\left[\frac{\psi''(\sigma^{-2})}{\sigma^{9}}+\frac{3\psi'(\sigma^{-2})}{\sigma^7}-\frac{2}{\sigma^5}\right]}
                     {\sqrt{n}\left[\frac{\psi'(\sigma^{-2})}{\sigma^6}-\sigma^{-4}\right]^{3/2}}$\\
Inver. Gaussian         &$\frac{8\sigma^6}{n^2}$&$\frac{2^{3/2}}{\sqrt{n}}$\\
\hline
\end{tabular}
\end{center}
\label{tabelaexpf2}
\end{table}

\subsection{Exponential Family Nonlinear Models}

We derive the skewness of the MLE $\hat\beta$ in EFNLMs. Under the
parametrization $t\{y,\tau^{-1}(\mu)\}= y\tau^{-1}(\mu)- b\{\tau^{-1}(\mu)\}$,
we have $d\tau^{-1}(\mu)/d\mu=V(\mu)^{-1}$, $d_2=-V^{-1}$, $d_3 = 2V^{-2}V^{(1)}$,
$W=\{V^{-1}(d\mu/d\eta)^2\}$ and the model matrix is $\widetilde{X}$.
Thus, equation (\ref{secskew}) reduces to
$$\kappa_3(\hat{\beta}_a)=\frac{1}{\phi^2} \sum_{i=1}^n \left[m_{ai}^{3} \left\{3\frac{d\mu}{d\eta}\frac{d^2\mu}{d\eta^2}V^{-1}- 2\left(\frac{d\mu}{d\eta}\right)^3 V^{-2}V^{(1)}\right\}_i - 3m_{ai}n_{ai}\left(\frac{d\mu}{d\eta}V^{-1}\right)_i\right],$$
where the matrix $N$ was defined in Section 2. The skewness of the
MLEs of $\phi$ and $\sigma^2$ are equal to those of Section 4.1,
since the nonlinearity does not affect these parameters. These 
results agree with those by Cavalcanti et al. (2009).

\subsection{Exponential Dispersion Models}

The skewness of the MLEs in EDMs has not been investigated and equation (\ref{secskew})
can be applied for several EDMs
discussed in J\o rgensen's (1997b) book, although the application of equation (\ref{kappasigma2})
is a much more difficult problem. For example, J\o rgensen (1997b)
discusses the Tweedie class of distributions with power variance function
defined by $V(\mu)=\mu^{\delta}$. The cumulant generator
function $b_{\delta}(\theta)$ for $\delta\neq1,2$ is
$$b_{\delta}(\theta) = (2-\delta)^{-1} \left\{(1-\delta)\theta \right\}^{\frac {\delta-2}
{\delta-1}},$$
and $b_{1}(\theta)=\exp(\theta)$ and $b_{2}(\theta)=-\log(-\theta)$.
We recognize for $\delta=0,2$ and $3$, the cumulant generator corresponding
to the normal, gamma and inverse Gaussian distributions, res\-pectively.
There exist continuous EDMs generated by extreme stable distributions with
support $\mathbb{R}$ and positive stable distributions for $\delta \leq 0$ and $\delta \geq 2$,
respectively, and compound Poisson distributions for $1<\delta<2$.
Setting $\alpha=(\delta-2)/(\delta-1)$, the function $a(\phi,y)$
for these two classes of models can be obtained from J\o rgensen (1997b).
For $\delta<0$ ($x\in \mathbb{R}$), we have
\begin{equation*}
a(\phi,y)=-\log\left(\pi y \right)
+ \log \left\{\sum_{j=1}^{\infty} m(j,\delta)(-y)^{j} \phi^{-j/(2-\delta)}\right\},
\end{equation*}
where
\begin{equation*}
m(j,\delta) = \frac{\Gamma(1+ j/\alpha)} {j!}
\left(\frac {\alpha} {\alpha-1}\right)^{j/\alpha}\sin \left(-j \pi/\alpha \right).
\end{equation*}
For $\delta >2$ ($y>0$), $a(\phi,y)$ is given by
\begin{equation*}
a(\phi,y) = - \log \left(\pi x \right)
 + \log \left\{\sum_{j=1}^{\infty} m(j,\delta) y^{-j}\phi^{-j}\right\},
\end{equation*}
where
\begin{equation*}
m(j,\delta)=\frac {\alpha \Gamma(1+ j \alpha)} {j!(\delta-1)}
\left\{\frac{(\delta-1)^\alpha}{(2-\delta)}\right\}^{j} \sin \left(-j \pi \alpha\right).
\end{equation*}
Our formulas do not depend on these complicated functions which are used only to
estimate $\phi$ for computing the skewness of the MLEs of $\beta$. 

We also would like to remark that there exists an exponential dispersion model with exponential
variance function, $V(\mu) = e^\mu$, for more details see the book of Jorgensen (1997b).

Table \ref{tabelaexpdisp} provides the basic quantities for the skewness
in generalized hyperbolic secant (GHS), negative binomial distributions, as well as for the skewness in the Tweedie distributions with power and exponential variance functions. These special cases have not been discussed in the literature so far.
The GHS distribution is defined by taking $b(\theta)=-\log\{\cos(\theta)\}$, 
whereas the term $a(\phi,y)$ in (\ref{density}) is given by
$$a(\phi,y)=\log\left\{\frac{2^{(1-2\phi)/\phi}}{\phi \Gamma(\phi^{-1})}\right\}
-\sum_{j=1}^{\infty} \log \left\{1+\frac {y^2} {(1+ 2 j \phi)^2}\right\}. $$
\begin{table}[hb]
\caption{Expressions for $d_2$, its derivative and $d_3$ for some EDMs.}
\begin{center}
\begin{tabular}{lccc}
\hline
Distribution & $d_2$&$d_2'$&$d_3$\\
\hline
GHS                      &$-\frac{2}{(\mu^2+1)^2}$ &$\frac{8\mu}{(\mu^2+1)^3}$& $\frac{(2\mu^3+10\mu)}{(\mu^2+1)^3}$\\
Neg. Bin.        &$-\frac{1}{\mu}+\frac{1}{1-\mu}$&$\left[\frac{1}{\mu^2}-\frac{1}{(1-\mu)^2}\right]$&$-\frac{2}{(1+\mu)^2}+\frac{2}{\mu^2}$\\
Power Var.           &$-\mu^{-p}$&$p\mu^{-(p+1)}$&$2p\mu^{-(p+1)}$\\
Exp. Var.     &$-e^{-\beta\mu}$&$\beta e^{-\beta\mu}$&$2\beta e^{-\beta\mu}$\\
\hline
\end{tabular}
\end{center}
\label{tabelaexpdisp}
\end{table}

\subsection{Proper Dispersion Models}

For PDMs, equation (\ref{secskew}) has no reduction, since the only difference between
PDMs and DMs is the form of the function $a(\cdot,\cdot)$, which can be decomposed
as $a(\phi,y)=a_1(\phi)+a_2(y)$. We now give the second-order third
cumulant of $\hat{\phi}$ and $\hat{\sigma}^2$. For PDMs, $\alpha^{(2)}=-n a_1''(\phi)$,
$\alpha_{3,0}=na_1'''(\phi)$, $\alpha^{(3)}=-na_1'''(\phi)$ and $\alpha_{2,1}=0$.
Using (\ref{kappaphi}), we have
$$\kappa_3(\hat{\phi})= -\frac{2 a_1'''(\phi)}{n^2 a_1''(\phi)^3}\,\,\text{and}\,\,
\gamma_1(\hat{\phi}) = \frac{2 a_1'''(\phi)}{\sqrt{n} \{-a_1''(\phi)\}^{3/2}}.$$
For $\hat{\sigma}^2$, we obtain
$$\kappa_3(\hat{\sigma}^2) = -\frac{2\sigma^4\{\sigma^2 \xi'''(\sigma^2)+3\xi''(\sigma^2)\}}{n^2\{2\xi'(\sigma^2)+\sigma^2\xi''(\sigma^2)\}^3},$$
and
$$\gamma_1(\hat{\sigma}^2) = \frac{2\sigma\{\sigma^2 \xi'''(\sigma^2)+3\xi''(\sigma^2)\}}{\sqrt{n}\{-2\xi'(\sigma^2)-\sigma^2\xi''(\sigma^2)\}^3}.$$
The form of $a(\phi,y)$ for this case is different of that one for the two-parameter
exponential family models but the expressions for the third cumulant and
skewness of $\hat\phi$ and $\hat\sigma^2$ are identical.

We illustrate the idea on a particular example of PDM. We consider
the \emph{von Mises regression model} that is quite useful for modeling
circular data (see, Mardia (1972) and Fisher (1993)).
Here, the density function is given by
\begin{equation}\label{vonmises}
\pi(y;\mu,\phi) = \frac{1}{2\pi I_0(\phi)} \exp\{\phi\cos(y-\mu)\},
\end{equation}
where $-\pi<y\leq \pi$, $-\pi<\mu\leq \pi$, $\phi>0$, and $I_v$ denotes the modified Bessel function
of the first kind and order $v$ (see Abramowitz and Stegun, 1970, Eq. 9.6.1). The density (\ref{vonmises})
is symmetric around $y=\mu$ which is both the mode and the circular mean of the distribution.
Here, $\phi$ is a precision parameter in the sense that when it increases,
the density function (\ref{vonmises}) becomes more concentrated around $\mu$.
Clearly, the density fucntion (\ref{vonmises}) is a PDM, since
$t(y,\mu) = \cos(y-\mu)$ and $a_1(\phi) = \log\{I_0(\phi)\}$. We investigate
the skewness of the estimaate of $\beta$. We have $E\{\sin(Y-\mu)\} = 0$ and
$E[\{\cos(Y-\mu)\}^2] = 1-\phi^{-1}r(\phi)$, where $r(\phi) = I_1(\phi)/I_0(\phi)$.
These results yield $d_2 = -r(\phi)$ and $d_3=d_2'=0$. The matrix $W$ is
$W = {\rm diag}\{(d\mu/d\eta)^2 r(\phi)\}$ and we can obtain the inverse of the information matrix,
and the matrices $M$ and $N$. Further, $f=(d\mu/d\eta)(d^2\mu/d\eta^2)r(\phi)$,
$g = (d\mu/d\eta)(d^2\mu/d\eta^2)r(\phi)$ and $e = 0$.
Hence, formula (\ref{secskew}) yields
$$\kappa_3(\hat{\beta}_a) = -\frac{3}{\phi^2} \sum_{i=1}^n\left\{ m_{ai}^3\frac{d\mu}{d\eta}\frac{d^2\mu}{d\eta^2}r(\phi) - m_{ai}n_{ai}\left(\frac{d\mu}{d\eta}\right)_i^2r(\phi)\right\}.$$
If the link function is the identity function, i.e.
$\eta=\mu$, then $w=r(\phi)$ and $f=g=e=0$. For a linear von Mises
regression model with identity link function, $\kappa_3(\hat{\beta}_a)=0$.
For a nonlinear model, we obtain
$$\kappa_3(\hat{\beta}_a) = \frac{3r(\phi)}{\phi^2} \sum_{i=1}^n m_{ai}n_{ai}.$$
First, for the skewness of the MLEs of $\phi$ and $\sigma^2$, we have
$I_0'(\phi)=I_1(\phi)$ and $I_1'(\phi) = I_0(\phi) - I_1(\phi)/\phi$ (Abramowitz and Stegun,
1970; equations 9.6.26 and 9.6.27). Then, $a_1''(\phi) = r'(\phi)$
and $a_1'''(\phi) = r''(\phi)$, where $r(\phi)=I_1(\phi)/I_0(\phi)$ as before.
Hence, we obtain for the von Mises model
$$\kappa_3(\hat{\phi})=-\frac{2 r''(\phi)}{n^2 r'(\phi)^3}\,\,\text{and}\,\,
\gamma_1(\hat{\phi})=\frac{2 r''(\phi)}{\sqrt{n} \{-r'(\phi)\}^{3/2}}.$$
Now, some similar calculations yield
$$\kappa_3(\hat{\sigma}^2) = \frac{2\sigma^{12}\left\{ 3\sigma^2 r'(\sigma^{-2}) + r''(\sigma^{-2})\right\}}{n^2 \{r'(\sigma^{-2})\}^3},$$
and then
$$\gamma_1(\hat{\sigma}^2) = \frac{2\left\{-3\sigma^{2} r'(\sigma^{-2}) - r''(\sigma^{-2})\right\}}{\sigma^6\sqrt{n}\{-r'(\sigma^{-2})\}^{3/2}}.$$
Table \ref{tabelaproper} lists the quantities required for several PDMs,
whereas Table \ref{tabelaproper2} gives the
skewness of the MLEs of $\phi$ and $\sigma^2$ for some PDMs.

\begin{table}[!h]
\caption{Expressions of $d_2$, its derivative and $d_3$ in PDMs.}
\begin{center}
\begin{tabular}{lccc}
\hline
Distribution & $d_2$&$d_2'$&$d_3$\\
\hline
Rec. Gamma         &$-\mu^{-2}$&$2\mu^{-3}$&$2\mu^{-3}$\\
Log-Gamma                &$-1$&$0$&$1$\\
Rec. Inv. Gauss. &$-\mu^{-1}$&$\mu^{-2}$&$0$\\
Von-Mises                &$-r(\phi)$&$0$&$0$\\
\hline
\end{tabular}
\end{center}
\label{tabelaproper}
\end{table}

\begin{table}[!h]
\caption{Skewness of $\hat{\phi}$ and $\hat{\sigma}^2$ for some PDMs.}
\begin{center}
\begin{tabular}{lcc}
\hline
Distribution&$\kappa_3(\hat{\phi})$&$\gamma_1(\hat\phi)$\\
\hline
Rec. Gamma          &$\frac{2\phi(1+\phi^2\psi''(\phi)]}{n^2[1-\phi \psi'(\phi)]^3}$&
                     $\frac{-2[\psi''(\phi)+\phi^{-2}]}{\sqrt{n}[\psi'(\phi)-\phi^{-1}]^{3/2}}$\\
Rec. Inv. Gauss.    &$\frac{16\phi^3}{n^2}$&$\frac{2^{5/2}}{\sqrt{n}}$\\
Log-Gamma           &$\frac{2\phi(1+\phi^2\psi''(\phi)]}{n^2[1-\phi \psi'(\phi)]^3}$&
                     $\frac{-2[\psi''(\phi)+\phi^{-2}]}{\sqrt{n}[\psi'(\phi)-\phi^{-1}]^{3/2}}$\\
von-Mises           &$\frac{-2r''(\phi)}{n^2 r'(\phi)^3}$&$\frac{-2r''(\phi)}{\sqrt{n}[r'(\phi)]^{3/2}}$\\
\hline
&$\kappa_3(\hat{\sigma}^2)$&$\gamma_1(\hat{\sigma}^2)$\\
\hline
Rec. Gamma&                    $\frac{-2\left[\frac{\psi''(\sigma^{-2})}{\sigma^{6}}+\frac{3\psi'(\sigma^{-2})}{\sigma^4}-\frac{2}{\sigma^2}\right]}
                      {n^2\left[\sigma^{-4}-\frac{\psi'(\sigma^{-2})}{\sigma^6}\right]^3}$&
                     $\frac{2\left[\frac{\psi''(\sigma^{-2})}{\sigma^{9}}+\frac{3\psi'(\sigma^{-2})}{\sigma^7}-\frac{2}{\sigma^5}\right]}
                      {\sqrt{n}\left[\frac{\psi'(\sigma^{-2})}{\sigma^6}-\sigma^{-4}\right]^{3/2}}$\\
Rec. Inv. Gauss.  &$\frac{8\sigma^6}{n^2}$&$\frac{2^{3/2}}{\sqrt{n}}$\\
Log-Gamma &
                     $\frac{-2\left[\frac{\psi''(\sigma^{-2})}{\sigma^{6}}+\frac{3\psi'(\sigma^{-2})}{\sigma^4}-\frac{2}{\sigma^2}\right]}
                      {n^2\left[\sigma^{-4}-\frac{\psi'(\sigma^{-2})}{\sigma^6}\right]^3}$&
                     $\frac{2\left[\frac{\psi''(\sigma^{-2})}{\sigma^{9}}+\frac{3\psi'(\sigma^{-2})}{\sigma^7}-\frac{2}{\sigma^5}\right]}
                      {\sqrt{n}\left[\frac{\psi'(\sigma^{-2})}{\sigma^6}-\sigma^{-4}\right]^{3/2}}$\\
von-Mises &
                     $\frac{2\sigma^{12}[r'(\sigma^{-2})\sigma^2+r''(\sigma^{-2})]}{n^2[r'(\sigma^{-2})]^3}$&
                     $\frac{2[r'(\sigma^{-2})\sigma^2+r''(\sigma^{-2})]}{\sqrt{n}[r'(\sigma^{-2})]^{3/2}}$\\
\hline

\end{tabular}
\end{center}
\label{tabelaproper2}
\end{table}

\newpage
\subsection{Some Other Special Submodels}

We investigate some special cases which were first studied by Cordeiro (1985).
If we take $t(y,\theta)= y\mu-b(\mu)$, (\ref{density}) is a one parameter exponential family
indexed by the canonical parameter $\mu$. Now, we assume that $t(y,\mu)$
involves a known constant parameter $c$ for all observations, say $t(y,\mu) = t(y,\mu,c)$, and
that $\phi=1$ and $a(\phi,y)= a(c,y)$. Several models can be defined in this framework:
normal $N(\mu,c^2\mu^2)$, log-normal $LN(\mu,c^2\mu^2)$ and inverse Gaussian $IG(\mu,c^2\mu^2)$
distributions with mean $\mu$ and known constant coefficient of variation $c$ and Weibull $W(\mu,c)$
distribution with mean $\mu$ and known constant shape parameter $c$. Here, the normal and inverse
Gaussian distributions are not standard GLMs since we consider a different parametrization.

For these models, we have $d_2= -k_2\mu^{-2}$, $d_3=k_3 \mu^{-3}$ and $d_2'= 2k_2\mu^{-3}$,
where $k_2$ and $k_3$ are known positive functions of $c$ (see Table \ref{tabelaspecial}).
The matrix $W$ becomes $W={\rm diag}\{k_2\mu^{-2}(d\mu/d\eta)^2\}$ and we can obtain the
inverse of the information matrix and the matrices $M$ and $N$. Further, $w=k_2\mu^{-2}(d\mu/d\eta)^2$,
$f=k_2\mu^{-2}(d\mu/d\eta)(d^2\mu/d\eta^2)-k_3\mu^{-3}(d\mu/d\eta)^3$, $g=k_2\mu^{-2}(d\mu/d\eta)(d^2\mu/d\eta^2)$
and $e=-2k_2\mu^{-3}(d\mu/d\eta)^3$. Then, equation (\ref{secskew}) yields
$$\kappa_3(\hat{\beta}_a) = \sum_{i=1}^n\left[m_{ai}\left\{(6k_2-k_3)\mu^{-3}\left(\frac{d\mu}{d\eta}\right)^3-3k_2\mu^{-2}\frac{d\mu}{d\eta}\frac{d^2\mu}{d\eta^2}\right\}_i-3m_{ai}n_{ai}k_2\left(\mu^{-2}\frac{d\mu}{d\eta}\right)_i^2\right]. $$
\begin{table}[htb]
\caption{Expressions of $k_2$ and $k_3$ for the normal, inverse Gaussian, log-normal and Weibull distributions.}
\begin{center}
\begin{tabular}{ccc}
\hline
Model            &$k_2$      & $k_3$\\
\hline
Normal ($N(\mu,c^2\mu^2)$)   &$c^{-2}(1+2c^2)$         & $c^{-2}(6+10c^2)$ \\
Inverse Gaussian ($IG(\mu,c^2\mu^2)$) &$1/2 c^{-2}(1+c^2)$ & $c^{-2}(3+c^2)$  \\
Log-normal ($LN(\mu, c^2\mu^2)$)& $[\log(1+c^2)]^{-1}$ & $3[\log(1+c^2)]^{-1}$ \\
Weibull ($W(\mu,c)$) & $c^2$ & $c^2(c+3)$ \\
\hline
\end{tabular}
\end{center}
\label{tabelaspecial}
\end{table}

\section{Simulation results}

We present some simulation results for the finite-sample distributions
of the skewness of the MLEs of $\beta$, $\phi$ and $\sigma^2$. We use a reciprocal
gamma model with square root link
$$\sqrt{\mu_i} = \beta_0 + \beta_1 x_{1,i} + x_{2,i}^{\beta_2},\qquad i=1,\ldots,n,$$
where the true values of the parameters were taken as $\beta_0 =
1/2$, $\beta_1 = 1$, $\beta_2=2$ and $\phi=4$. The elements of the $n\times 3$ matrix $\widetilde{X}$ are: $\widetilde{X}_{i,1}=1$; $\widetilde{X}_{i,2}=x_{1,i}$; and $\widetilde{X}_{i,3}=\log(x_{2,i})x_{2,i}^{\beta_2}$. The explanatory
variables $x_1$ and $x_2$ were generated from the uniform U$(0,1)$ and $U(1,2)$
distributions, respectively, for $n=20, 40$ and $60$. The values of $x_1$ and $x_2$ were held constant
throughout the simulations. The number of Monte Carlo replications
was set at $10,000$ and all simulations were performed using the
statistical software {\tt R}.

In each of the $10,000$ replications, we fitted the model and computed the
MLEs $\hat\beta_0$, $\hat\beta_1$, $\hat\beta_2$, the fitted values $\hat\mu_1,\ldots,\hat\mu_n$, $\hat\phi$ and $\hat\sigma^2$. Then, we computed their estimated asymptotic skewness $\hat\gamma_1(\hat\beta_0), \hat\gamma_1(\hat\beta_1), \hat\gamma_1(\hat\beta_2), \hat\gamma_1(\hat\phi)$ and $\hat\gamma_1(\hat\sigma^2)$, where each unknown value is replaced by its MLE, and their true asymptotic skewness $\gamma_1(\hat\beta_0), \gamma_1(\hat\beta_1), \gamma_1(\hat\beta_2), \gamma_1(\hat\phi)$ and $\gamma_1(\hat\sigma^2)$. By true asymptotic skewness we mean the asymptotic skewness calculated by using the true values of the regression parameters. We then computed the sample skewness $\hat g_3(\hat\beta_0), \hat g_3(\hat\beta_1), \hat g_3(\hat\beta_2), \hat g_3(\hat\phi)$ and $\hat g_3(\hat\sigma^2)$, where $\hat g_3(\cdot)$ is given by $\hat g_3(\alpha) = m_3(\alpha)/m_2(\alpha)^{3/2}$, for a scalar $\alpha$, and $m_r(\alpha) = \sum_{i=1}^{10000} (\alpha_i - \overline{\alpha})^r$ and
$\overline{\alpha} = \frac{1}{10000}\sum_{i=1}^{10000}\alpha_i$.

Table \ref{tabelabeta} gives the sample means of the estimated skewness $\hat\gamma_1(\hat\beta_0), \hat\gamma_1(\hat\beta_1)$ and $\hat\gamma_2(\hat\beta_2)$, the true skewness $\gamma_1(\hat\beta_0), \gamma_1(\hat\beta_1)$ and $\gamma_1(\hat\beta_2)$, and the sample skewness $\hat g_3(\hat\beta_0), \hat g_3(\hat\beta_1)$ and $g_3(\hat\beta_2)$.

\begin{table}[h!]
\caption{Estimated, true and sample skewness of $\hat\beta_0$, $\hat\beta_1$ and $\hat\beta_2$}
\begin{center}
\begin{tabular}{lccccccccc}
\hline
n & $\hat\gamma_1(\hat\beta_0)$&$\gamma_1(\hat\beta_0)$&$\hat g_3(\hat\beta_0)$& $\hat\gamma_1(\hat\beta_1)$&$\gamma_1(\hat\beta_1)$&$\hat g_3(\hat\beta_1)$& $\hat\gamma_1(\hat\beta_2)$&$\gamma_1(\hat\beta_2)$&$\hat g_3(\hat\beta_2)$\\
\hline
20 & -0.1132 & -0.1355 & -0.6519 & -0.0454 & -0.0753 & -0.1408 & 1.2430 & 2.4222 & 7.9249\\
40 & -0.0989 & -0.1121 & -0.3233 & -0.0365 & -0.0577 & -0.1282 & 0.9846 & 1.6590 & 3.5451\\
60 & -0.0545 & -0.0870 & -0.1684 & -0.0184 & -0.0405 & -0.0997 & 0.5673 & 1.1399 & 2.1091\\
\hline
\end{tabular}
\end{center}
\label{tabelabeta}
\end{table}

The figures in Table \ref{tabelabeta} show that the sample and analytical skewness
decrease as the sample size increases, in agreement with the first-order asymptotic
theory. We also note that $\hat{\beta_0}$ and $\hat\beta_1$ are always negatively skewed, whereas $\hat\beta_3$ is always positively skewed. In most of the cases, the sample skewness larger, in absolute values,
than the estimated asymptotic skewness. We note that the estimated and true asymptotic skewness
are not far apart. We observe large differences between the sample skewness and the estimated asymptotic 
skewness for $n=20$. The explanation for such behavior is that the expected value of $m_r$ is equal to the $r$th central moment of the population if we neglect terms of order $n^{-1/2}$. These terms, however, are not negligible for small sample sizes.

Table \ref{tabelaphi} gives the sample mean of the estimated asymptotic skewness $\hat\gamma_1(\hat\phi)$ and $\hat\gamma_1(\hat\sigma^2)$ out of $10,000$ values, the true asymptotic skewness $\gamma_1(\hat\phi)$ and $\gamma_1(\hat\sigma^2)$, and the sample skewness, $\hat g_3(\hat\phi)$ and $\hat g_3(\hat\sigma^2)$.

\begin{table}[h!]
\caption{Estimated, true and sample skewness of $\hat\phi$ and $\hat\sigma^2$.}
\begin{center}
\begin{tabular}{lcccccc}
\hline
n & $\hat\gamma_1(\hat\phi)$&$\gamma_1(\hat\phi)$&$\hat g_3(\hat\phi)$& $\hat\gamma_1(\hat\sigma^2)$&$\gamma_1(\hat\sigma^2)$&$\hat g_3(\hat\sigma^2)$\\
\hline
20 & 0.8732 & 1.0911  & 1.8858& 0.3154 & 0.4878 & 0.6998\\
40 & 0.6493 & 0.7809 & 1.0072& 0.2682 & 0.3449 & 0.4364\\
60 & 0.3927 & 0.5443 & 0.8003 & 0.2319 & 0.2816 & 0.2922\\
\hline
\end{tabular}
\end{center}
\label{tabelaphi}
\end{table}

From the figures in Table \ref{tabelaphi}, it is clear that the asymptotic normality of $\hat\phi$ and $\hat\sigma^2$, often used in DMs is not achieved for small values of $n$. The results in this table suggest that there
is a quite reasonable agreement between the analytical and the sample skewness. The estimated and
true skewness are quite close even for small values of $n$.

\section{Conclusion}

In in this article, we introduce the dispersion models (DMs) with a regression systematic
component to extend the well-known generalized linear models (GLMs), the exponential family
nonlinear models (EFNLMs) (Cordeiro and Paula, 1989) and the class of proper dispersion models
(PDMs) (J\o rgensen, 1997a). Several properties of distributions in the class of DMs are
discussed in the excellent book of J\o rgensen (1997b). For the first time, we derive the
second-order skewness of the MLEs of the regression parameters in DMs using formulae
obtained by Bowman and Shenton (1998).

We obtain an explicit matrix expression for the skewness of the maximum likelihood
estimate (MLE) of the regression parameter vector $\beta$. We also derive the skewness
of the MLEs of the precision and dispersion parameters. Our results generalize those
obtained by Cordeiro and Cordeiro (2001) and Cavalcanti et al. (2009), and also provide 
new results for some special submodels such as the exponential dispersion 
models and PDMs. In particular, we discuss results for the Von-Mises regression model. 
We perform a simulation study in a nonlinear reciprocal gamma model that indicates that the 
normal approximation usually employed with MLEs in DMs can be misleading in 
samples with small to moderate sizes.


\begin{thebibliography}{30}
\bibitem{14} Abramowitz, W. and Stegun, I.A. (1970). \emph{Handbook of Mathematical Functions with Formulas,
Graphs and Mathematical Tables.} Washington, National Bureau of Standards. (National Bureau of Standards. Applied Mathematics Series, 55).
\bibitem{10} Bowman, K.O. and Shenton, L.R. (1998). Asymptotic skewness and the distribution of maximum likelihood estimators. \emph{Comm. Statist. Theory Meth.,} 27, 2743-2760.
\bibitem{11} Cavalcanti, A.B., Cordeiro, G.M., Botter, D.A. and Barroso, L.P. (2009).
Asymptotic skewness in exponential family nonlinear models. \emph{Comm. Statist. Theory Meth.} To appear.
\bibitem{15} Cordeiro, G.M. (1985). The null expected deviance for an extended class of
gene\-ralized linear models. \emph{Lecture Notes in Statistics}, 32, 27-34.
\bibitem{6} Cordeiro, G.M. and McCullagh, P. (1991). Bias correction in
generalized linear models. \emph{J. Roy. Statist. Soc. B,} 53, 629-643.
\bibitem{4} Cordeiro, G.M. and Paula, G.A. (1989). Improved likelihood ratio statistic for exponential family nonlinear models. \emph{Biometrika,} 76, 93-100.
\bibitem{7} Cordeiro, G.M., Paula, G.A. and Botter, D.A. (1994). Improved likelihood ratio tests
for dispersion models. \emph{Inter. Statist. Rev.,} 62, 257-274.
\bibitem{111} Cordeiro, H.H. and Cordeiro, G.M. (2001). Skewness for parameters in generalized linear models. \emph{Comm. Statist. Theory Meth.}, 30, 1317-1334.
\bibitem{3} Cox, D.R. and Hinkley, D.V. (1974). \emph{Theoretical Statistics.} London, Chapman and Hall.
\bibitem{8} Fisher, N.I. (1993). \emph{Statistical Analysis of Circular Data.} New York, Cambridge University
Press.
\bibitem{19} J\o rgensen, B. (1987a). Exponential dispersion models (with discussion). \emph{J. Roy. Statist. Soc., Ser. B,} 49, 127-162.
\bibitem{25} J\o rgensen, B. (1987b). Small-dispersion asymptotics. \emph{Brazilian J. Probab. Statist.,} 1, 59-90.
\bibitem{1} J\o rgensen, B. (1997a). Proper dispersion models (with discussion). \emph{Brazilian J. Probab. Statist.}, 11, 89-140.
\bibitem{2} J\o rgensen, B. (1997b). \emph{The Theory of Dispersion Models}. London, Chapman and Hall.
\bibitem{9} Mardia, K.V. (1972). \emph{Statistics of directional data.} Academic Press, New York.
\bibitem{5} Paula, G.A. (1992). Bias correction for exponential family nonlinear models.
\emph{J. Statist. Comput. Simul.}, 40, 43-54.
\bibitem{18} Rocha, A. V., Simas, A. B. and Cordeiro, G. M. (2009). Second-order asymptotic expressions for the covariance matrix of
maximum likelihood estimators in dispersion models. \emph{Stat. Prob. Let.} To appear.
\bibitem{17} Simas, A.B., Barreto-Souza, W. and Rocha, A.V. (2009a). Improved estimators for dispersion models with dispersion covariates. \emph{Submitted.}
\bibitem{12} Simas, A.B. and Cordeiro, G.M. (2009). Adjusted Pearson residuals in exponential family nonlinear models. \emph{J. Statist. Comput. Simul.,} 79, 411-425.
\bibitem{16} Simas, A.B., Cordeiro, G.M. and Nadarajah, S. (2009b). Asymptotic tail properties of the distributions in the class of dispersion models. \emph{Preprint:} arXiv:0809.1840
\bibitem{13} Wei, B-C. (1998). \emph{Exponential Family Nonlinear Models}. Singapore, Springer.
\end{thebibliography}
\end{document}